\documentclass{iopjournal}
\usepackage{setspace}

\begin{document}

\articletype{Article type} 

\title{Kovacs-like memory effect in strain stiffening collagen networks}

\author{Abhishek Ghadai$^1$\orcid{0009-0004-0413-8274} and Sayantan Majumdar$^{1,*}$\orcid{0000-0003-3431-2128}}

\affil{$^1$Soft Condensed Matter Group, Raman Research Institute, Bengaluru 560080, India}


\affil{$^*$Author to whom any correspondence should be addressed.}

\email{smajumdar@rri.res.in}


\begin{abstract}

Materials driven far from equilibrium can encode memories of past deformations through long-lived structural reorganisations. Such memory effects-reflecting parameters such as deformation direction, magnitude, and duration have been widely explored in soft amorphous solids. Here, we report a Kovacs-like memory effect manifested as a non-monotonic stress relaxation in vitro biopolymer networks formed by collagen, an essential component of the mammalian extracellular matrix. Using shear rheology combined with in-situ optical imaging, we find that this memory effect emerges exclusively in the nonlinear strain-stiffening regime, and persists over a much broader range of strain amplitudes than previously reported for other viscoelastic amorphous materials. Furthermore, we uncover a strong correlation between the memory response and the development of negative normal stresses and associated strain fields, highlighting the unique nonequilibrium mechanics underlying memory formation in biopolymer networks.
\end{abstract}

\section{Introduction}
Memory formation refers to a system's ability to encode, preserve, and access information regarding its past states \cite{keim2019memory}. This phenomenon is closely associated with out-of-equilibrium dynamics, as systems in thermal equilibrium completely forget their previous states. Recent studies point out that the mechanical properties of soft, amorphous materials can be tuned beyond their usual linear and non-linear viscoelasticity by encoding mechanical memories \cite{chattopadhyay2022inter,mandal2025kovacs}. These observations suggest that understanding memory formation in soft, disordered materials opens up possibilities for designing multifunctional, smart, and adaptive materials.  
Materials can memorize different types of deformation parameters. These include memory of maximum deformation amplitude, direction, duration, return point memory, shape memory, etc. \cite{keim2019memory, paulsen2024mechanical}. However, predicting the type of memory that can be encoded in a disordered material from its structural parameters can be complex and often unclear.
\newline
\newline
The Kovacs effect is an interesting memory phenomenon in glassy materials, where the memory is characterised by a non-monotonic (non-steadily increasing or decreasing) relaxation to reveal the system's previous deformation history. The classic experiment described by Kovacs involves thermal history \cite{kovacs2006transition}. In that experiment, a glassy sample was quickly cooled below its glass transition temperature, held there for a specific time called the ``waiting time" ($t_w$), and then was instantly reheated to a slightly higher temperature. The system's volume exhibits a surprising non-monotonic evolution, rather than a simple monotonic approach towards its new equilibrium state \cite{prados2010kovacs}. The non-monotonicity in the volume often manifests as a ``peak" in the relaxation curve. A hallmark of the Kovacs effect is the linear correlation between the peak time $t_p$ ( the time at which the non-monotonic response reaches its maximum after the second perturbation ) and the waiting time $t_w$.
\newline
\newline
The Kovacs-like memory effect is observed not only using thermal perturbations but also by mechanical ones, such as under compression-decompression and volume-preserving shear in amorphous systems like crumpled sheets \cite{lahini2017nonmonotonic}, elastic foams \cite{lahini2017nonmonotonic}, colloidal glasses \cite{mandal2025kovacs}, and granular packings \cite{murphy2020memory}. The Kovacs effect under mechanical perturbations is seen not only for two-step protocols but also for three or multi-step protocols \cite{murphy2020memory}. 

Kovacs effect is primarily observed in systems with complex relaxation processes demonstrating logarithmic or stretched exponential relaxation. Non-monotonicity occurs because different structures or ``modes" of the system relax at varying rates. After a two-step perturbation, the ``slow" modes from the initial perturbed state take longer to adjust, while the ``fast" modes quickly adapt to the second perturbation. This creates a situation where the fast and slow modes relax in opposing directions, giving rise to a non-monotonic relaxation process. Eventually, at long time scales, the slower elements dominate, causing the stress evolution to shift in the opposite direction.

Surprisingly, despite the highly complex interplay between the relaxation modes,  linear response theory (LRT) provides a simple and robust framework to describe Kovacs memory for small deformations \cite{mandal2021memory}.
\newline
\newline
Although Kovacs effect under mechanical deformation has been studied for a range of amorphous, glassy materials in recent years, to the best of our knowledge, such effects have never been explored in the context of biopolymer networks that form an important class of disordered soft materials. Physiologically important biopolymer networks formed by protein filaments like actin, collagen, fibrin, etc., play an important role in determining the mechanical properties of mammalian cells or, extracellular matrices. Due to the semi-flexible nature of biopolymers, these networks show very interesting non-linear mechanics like non-linear strain stiffening and negative normal stresses. Such non-linear mechanics have been studied widely using experiments, simulations and analytical models \cite{storm2005nonlinear,stein2011micromechanics,piechocka2016multi,janmey2007negative,broedersz2014modeling}.
Entangled biopolymer networks formed by fairly low monomer concentrations are structurally quite different from the conventional amorphous and glassy solids. Remarkably, despite significant structural differences, biopolymer networks also display highly non-exponential, glassy relaxation \cite{yang2022microscale,nam2016strain}. 
\newline
\newline
In the present study, we investigate a Kovacs-like memory effect in a disordered, entangled network formed by type-I collagen. Our experiments reveal a pronounced non-monotonic stress relaxation response characterized by a robust Kovacs peak and a strong linear correlation between the peak time and the waiting time. Interestingly, this effect is observed exclusively within the non-linear strain-stiffening regime. Moreover, the Kovacs peak persists over an exceptionally broad range of deformation parameters, significantly exceeding those reported in previous studies. Through a combination of rheological measurements and in-situ boundary imaging, we elucidate that the cooperative interplay between non-linear strain stiffening and normal stress effects underlies the emergence of the Kovacs effect over such an extended parameter space.

\section{Materials and methods}
We use rat tail type-I collagen monomer solution [4 mg/ml stock solution] purchased from Gibco, United States. The collagen monomers of specific concentrations are polymerized with 10X PBS solution, 1N NaOH solution, and deionized water [Publication number MAN0007327]. We fluorescently label the collagen fibers and use confocal microscopy to characterize the network architecture \cite{adam2023network}. For the rheological measurements, we use a MCR-702 stress-controlled rheometer (Anton-Paar, Graz, Austria) using cone-plate geometry (diameter = $25\, mm$, cone angle $=2^{\circ}$). To obtain maximal adhesion of collagen polymers to the rheometer plates, we treat the rheometer plates with poly-L-lysine solution. For the coating process, we first thoroughly clean the rheometer plates using a soap solution, followed by rinsing with ethanol and deionized water. Next, we deposit $200\mu l$ of a 0.001\% poly-L-lysine solution (Sigma-Aldrich) onto the plates and incubate them at room temperature for 10 minutes. Afterward, we eliminate any excess solution by rinsing the plates multiple times with ethanol and deionised water, then let them dry.
We perform in-situ imaging at the flow-gradient plane of the sample boundary in the cone-plate setup. To map the local flow field, the sample is seeded with polystyrene tracer beads of average diameter of $3.3 \,\mu m$ and the images are analyzed using particle imaging velocimetry by PIV Lab\cite{Thielicke_2021} using MATLAB software.

\section{Results and discussion}

\begin{figure}
 \centering
        \includegraphics[width=1\textwidth]{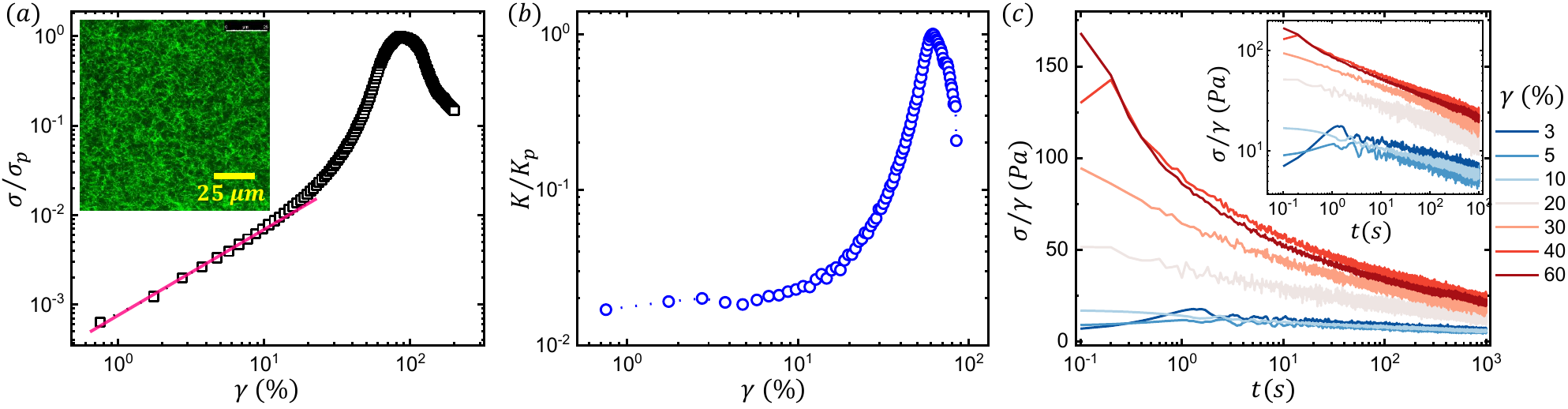}
 \caption{(a)Stress-strain curve for 2 mg/ml collagen polymerized at $25^{\circ} C$. The linear relation between stress and strain is marked as the red line. Confocal microscopy image of the network is shown in the inset. (b) Differential shear modulus as a function of shear strain. The differential shear modulus remains constant in the linear response region and increases substantially in the non-linear region. (c) Relaxation curves for different strain values corresponding to the linear and the strain stiffening regions. The shear modulus ($\sigma/\gamma$) is plotted w.r.t. time for various $\gamma$ values. The relaxation rate is higher for applied strains lying in the strain stiffening region compared to the linear region.}
\label{fig1}
\end{figure}

We first polymerize the collagen monomer into an entangled network. \textcolor{black}{For this, we mix desired amount of collagen monomers with 10X PBS, 1N NaOH and de-ionised water on ice and immediately deposit the mixture on a peltier temperature controlled  plate. The Peltier temperature is then raised to a desired value to initiate polymerization process. To prevent solvent evaporation, we deposit a thin layer of 5cSt silicon oil (Merck) and increase the humidity using a customised humidity chamber.} Once the polymerization is complete, the monomers form a system-spanning entangled network of collagen fibers and bundles. Using confocal microscopy, we directly visualize this structure for monomer concentration of 2 mg/ml polymerized at $25^{\circ}C$, as shown in Figure \ref{fig1}a (inset). We further quantify the network structure in terms of the average mesh size ($\zeta_{av}$). To calculate $\zeta_{av}$, we plot the distribution of distances along a line (taken both horizontally and vertically) between two consecutive bright pixels of a binary image. Before binarization, we employ a pixel classification method using Ilastik software \cite{berg2019ilastik} to reduce background noise in the confocal images. The distribution typically shows an exponential decay, which has also been observed in the earlier studies \cite{KAUFMAN,mesh}. The distribution is fitted with an exponential decay, and the average mesh size (\(\zeta_{av}\)) is defined as the decay exponent. For the network with monomer concentrations of 2 mg/ml with $25^{\circ}C$ polymerization temperature, the calculated value of \(\zeta_{av}\) is $2.1\,\pm 0.4 \,\mu m$.

To monitor the polymerization process during rheology, we measure the time evolution of the storage modulus by applying a sinusoidal strain of smaller magnitude (2\% at 0.5 Hz) over a sufficiently long time. Initially, the storage modulus value is low, but it begins to increase as the polymerization gradually forms an entangled network. Eventually, the storage modulus tend to saturate at longer times once the polymerization process is complete [Figure S1]. We identify the linear and non-linear regions of the polymerized collagen network by applying steady shear with a strain rate of \(0.01\, s^{-1}\) (where strain increases by $1\%$ each second). Figure \ref{fig1}a illustrates the stress-strain relationship of a 2 mg/ml collagen network polymerized at \(25^{\circ}C\). In this figure, the stress value at each applied strain is normalized with respect to the peak or the maximum stress value ($\sigma_p$) obtained over the entire measurement. The normalized stress $({\sigma}/{\sigma_p})$ varies linearly with strain \(\gamma\) up to \(\gamma \approx 10\%\). Beyond \(\gamma = 10\%\), ${\sigma}/{\sigma_p}$ starts to exhibit non-linear behaviour, demonstrating strain stiffening, where the rate of change of stress increases more rapidly compared to the rate of change of strain. We quantify such a rate of change of stress as the differential shear modulus $K=\frac{d\sigma}{d\gamma}$, and it is plotted in Figure \ref{fig1}b. We find that the value of $K/K_p$ where, $K_p = [\frac{d\sigma}{d\gamma}]_{max}$ increases rapidly with applied strain beyond the linear region, demonstrating a non-linear strain-stiffening response. The phenomenon of strain stiffening is also observed in a wide range of semi-flexible biopolymer networks \cite{storm2005nonlinear, majumdar2018mechanical}. The strain stiffening in semiflexible biopolymers has also been captured in various theoretical models\cite{broedersz2014modeling}. 
Non-linear strain stiffening not only alters the mechanical properties of biopolymer networks but also affects their relaxation dynamics. 
We perform stress relaxation measurement at various step strain magnitudes spanning the linear and non-linear strain stiffening regions. We obtain slow, non-exponential stress relaxation in all cases,  as shown in Figure \ref{fig1}c.
We also find from Figure \ref{fig1}c that the relaxation rate significantly increases for the applied strain values showing strain stiffening, compared to those in the linear region (see inset of Figure \ref{fig1}c). This is attributed to strain-enhanced stress relaxation, and also reported for collagen biopolymer networks earlier \cite{nam2016strain}.

\begin{figure}
 \centering
        \includegraphics[width=1\textwidth]{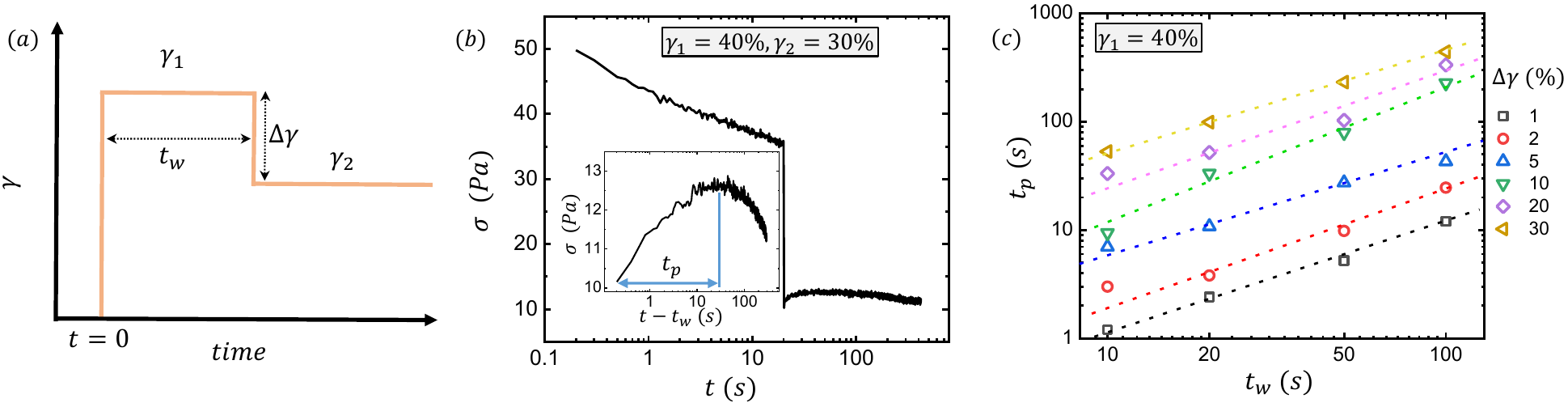}
 \caption{ (a) Schematic of the experimental protocol for the Kovacs memory effect. (b) Temporal evolution of stress showing Kovacs effect for $\gamma_1\,=\,40\%, \, \gamma_2 \,=\, 30\%, \, t_w\,=\, 20\,s$, and the inset shows the stress evolution during the second step $\gamma_2 = 30\%$. (c) Relation between the peak position of the stress response during the second step strain perturbation and the waiting time for $\gamma_1 \, = \, 40\%$ for different values of $\Delta \gamma$.  }
\label{fig2}
\end{figure}

Next, we explore the Kovacs-like memory effect in collagen networks. Figure \ref{fig2}a illustrates the schematic of the experimental protocol for a two-step Kovacs memory experiment. On a fully polymerized collagen network, we first apply a step shear strain $\gamma_1$ to the sample, and hold that strain for a specific time duration called the waiting time $t_w$. Next, the system's strain state is set to a new value $\gamma_2$, where $\gamma_2 < \gamma_1$ and $\Delta \gamma = \gamma_1 - \gamma_2 > 0$. The system is held at $\gamma_2$ for the rest of the experiment, and the time evolution of the shear stress is recorded continuously. Figure \ref{fig2}b illustrates the typical stress response of the system. Right after the application of the initial strain $\gamma_1$, there is a rapid increase in stress due to the elastic response of the system. Subsequently, the stress relaxes roughly logarithmically. When the strain state of the system is suddenly altered to $\gamma_2$, in the opposite direction of the previously applied $\gamma_1$, there is another sudden drop in stress because of the elastic response, followed by further relaxation of the system. From the stress response of the system after the application of $\gamma_2$ (also shown in the inset of Figure \ref{fig2}b), we find a non-monotonic evolution of stress in the system. Such non-monotonicity in stress clearly indicates Kovacs-like memory formation in the system. The time at which the peak in stress occurs after the second step $\gamma_2$ is referred to as the peak time $t_p$.
We perform experiments over a broad range of $\gamma_1$ and $\gamma_2$ values that cover both the linear and non-linear (strain stiffening) region of the system [Figure \ref{fig1}]. We also varied the waiting times $t_w$, ranging from 10 seconds to 100 seconds. Interestingly, we observe non-monotonic relaxation only for $\gamma_1$ values in the non-linear strain stiffening regime. Although the signal-to-noise level in the stress relaxation data is low for $\gamma_1$ values in the linear regime due to the small value of the stress, but average relaxation data does not demonstrate any non-monotonic evolution as shown in Figure S2. Nonetheless, over a broad range of $t_w$ and $\Delta \gamma$ values, we observe a clear peak in stress relaxation, indicating a non-monotonic stress evolution for large values of $\gamma_1$ in the non-linear strain stiffening regime.
Figure \ref{fig2}c shows the variation of peak time ($t_p$) w.r.t. waiting time ($t_w$) for a range of $\Delta \gamma$ from 1\% to 30\% for $\gamma_1 = 40\%$. We obtain a linear variation of $t_p$ w.r.t. $t_w$ for all values of $\Delta \gamma$, indicating the system retains a memory of the waiting time. We also obtain robust Kovacs-like memory effects in the non-linear strain stiffening regime for different collagen concentrations [Figures S3 and S4]. The Kovacs memory reported in earlier work mostly deals with the linear response regime \cite{lahini2017nonmonotonic,murphy2020memory, mandal2021memory}. Recently, Kovacs-like memory has also been studied in detail in the nonlinear region for a colloidal glass of soft particles, where the role of non-affine deformations has been explored \cite{mandal2025kovacs}. However, in colloidal glass the $\Delta\gamma$ range over which non-monotonic stress relaxation is observed for large $\gamma_1$ values in the non-linear regime is very small. This is highlighted by a narrow range of $\frac{\Delta\gamma}{\gamma_1}$ values over which Kovacs effect is observed. In this system, the non-linear region originates from the yielding of the colloidal glass at large deformations. On the other hand, in biopolymer networks, the non-linear strain stiffening takes place well below the yield point. Therefore, non-monotonic stress relaxation and Kovacs memory in the strain-stiffening biopolymer networks present an exciting direction to explore.
\begin{figure}
 \centering
        \includegraphics[width=0.85\textwidth]{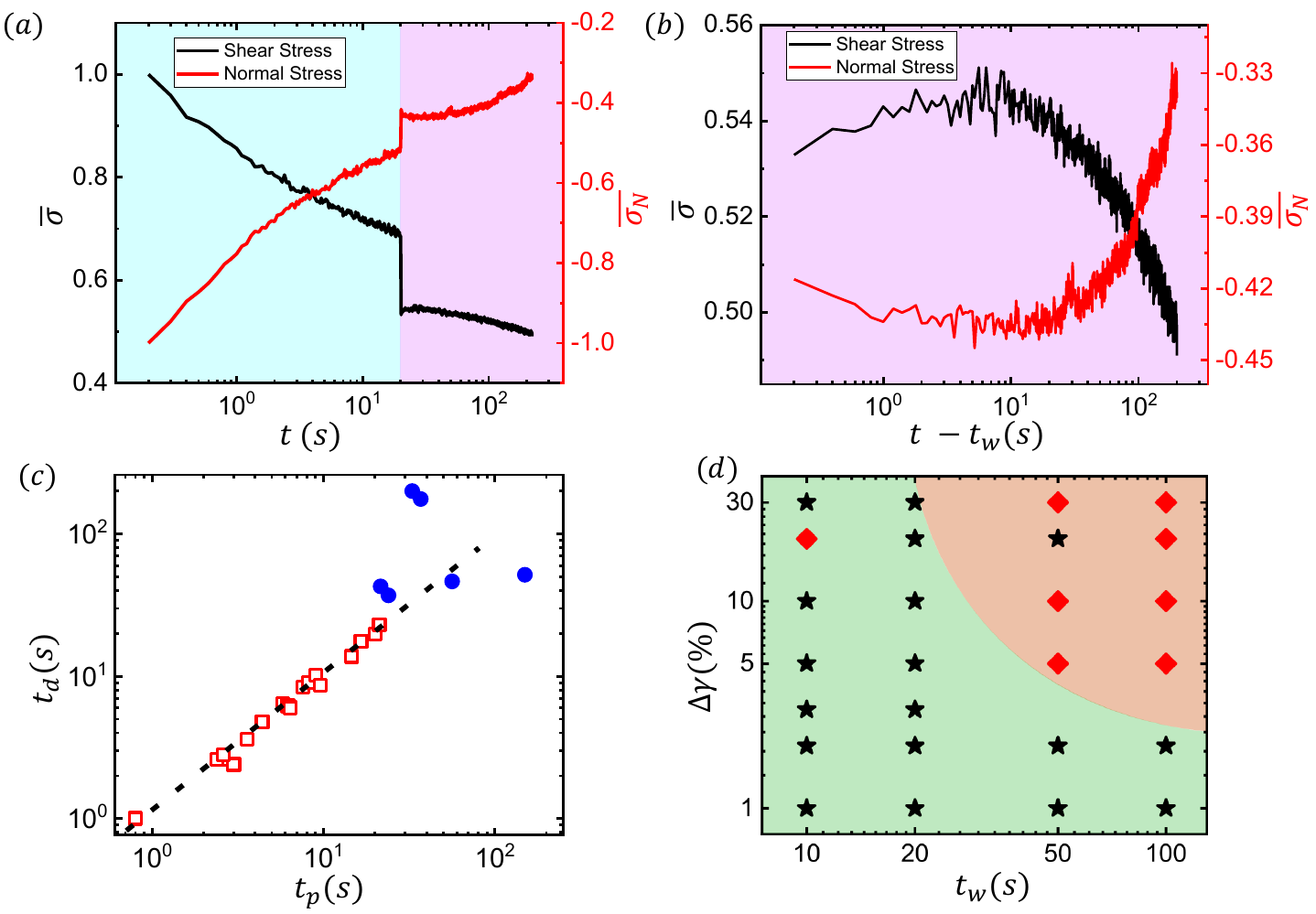}
 \caption{ (a) Time evolution of both shear and normal stresses for a two-step Kovacs memory protocol for $\gamma_1 = 40\%, t_w=20\,s$ and $\gamma_2=35\%$. Both the stresses are normalized with respect to the first point. Normalized shear and normal stresses only for the second step strain deformation are indicated in panel b. Both the peak time $t_p$ in shear stress and dip time $t_d$ in normal stress are close to each other. (c) peak time $t_p$ versus dip time $t_d$ showing good linear correlations, shown as red squares, and the deviations are shown as blue circles. (d) Phase diagram in the parameter plane of $\Delta \gamma$ and $t_w$ for $\gamma_1$ = 40\% for correlation between the time evolution of shear and normal stresses. The correlation points are marked as black stars and the anti correlation points are marked as red diamonds.}
\label{fig3}
\end{figure}
\newline
\newline
Semi-flexible biopolymer networks demonstrate negative normal stress under large deformations \cite{janmey2007negative}. For the collagen network, we find a very pronounced negative normal stress when strain stiffening occurs. In Figure \ref{fig3}a, we depict the time evolution of shear ($\sigma$) and normal stress ($\sigma_N$) during a typical two-step strain protocol discussed above for  $\gamma_1 = 40\%$, $t_w = 20\,s$, and  $\gamma_2 = 35\%$. The stress values are normalized with respect to the first point in Figure \ref{fig3}a. Figure \ref{fig3}b shows the time evolution of both stresses during the second step perturbation. We find that both shear and normal stresses exhibit non-monotonic behavior: the shear stress reaches a peak, while the normal stress shows a dip. Using a similar definition for peak time ($t_p$) in shear stress, we define a dip time ($t_d$) for normal stress as well. In Figure \ref{fig3}c, we plot $t_p$ versus $t_d$ for different values of $\gamma_1,\gamma_2$ and $t_w$. We find a strong linear correlation between $t_p$ and $t_d$ for a wide range of $\Delta \gamma$ and $t_w$ [Figure \ref{fig3}b]. However, this correlation does not hold (shown as blue circles) for larger $\Delta \gamma$ (\textcolor{black}{Figure S5}), even though the linearity between $t_p$ and $t_w$ remains valid. Furthermore, in some cases with longer waiting times, the evolution of normal stress becomes more complicated, showing either multiple non-monotonic variations or no dip at all, while the shear stress shows a clear non-monotonic behavior (\textcolor{black}{see Figure S5}). To investigate the dependence of the strength of correlation between the time evolution of shear and the normal stress,  we construct a phase diagram in the parameter plane of $\Delta\gamma$ and $t_w$ for $\gamma_1 = 40\%$  as shown in figure \ref{fig3}d. The points marked as black stars represent a one-to-one linear correlation, while and red diamonds represent poor or no correlation between $\sigma$ and $\sigma_N$. We find that a strong correlation holds over a wide range of waiting times when $\Delta \gamma$ values are small, and also over a large range of $\Delta \gamma$ values when the $t_w$ values are small. The decorrelation is observed otherwise.

\begin{figure}
 \centering
        \includegraphics[width=0.8\textwidth]{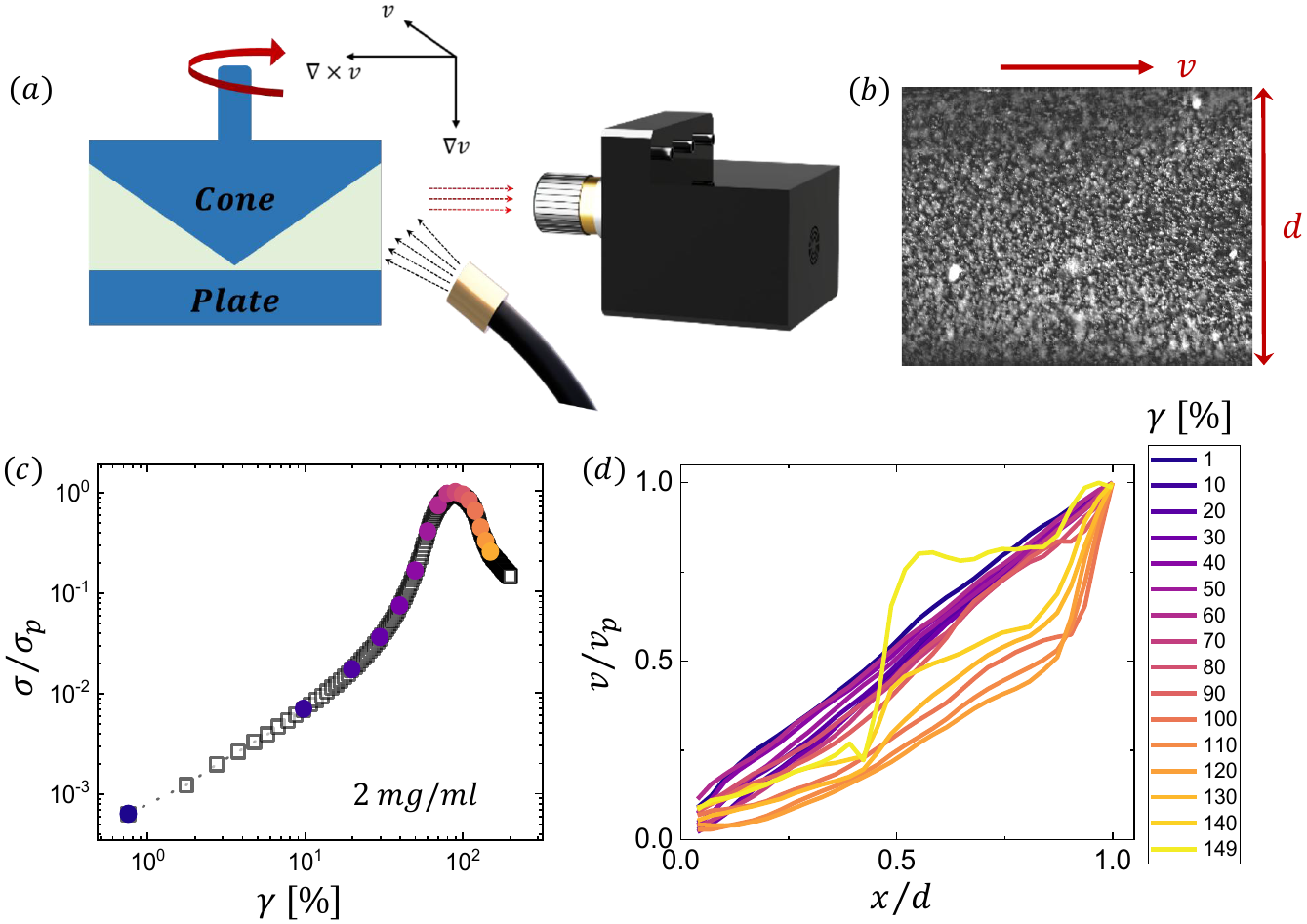}
 \caption{ (a) Schematic of the in-situ imaging setup. (b) Snapshot of the sample seeded with polystyrene tracer beads at the sample-air interface. (c) Stress-strain curve for 2 mg/ml collagen network seeded with polystyrene tracers polymerized at $25^{\circ}C$. (d) Normalized velocity($v/v_p$) as a function of normalized gap ($x/d$) for different strain values (indicated as the variation in the colorscale). }
\label{fig4}

\end{figure}

\begin{figure}
 \centering
        \includegraphics[width=0.9\textwidth]{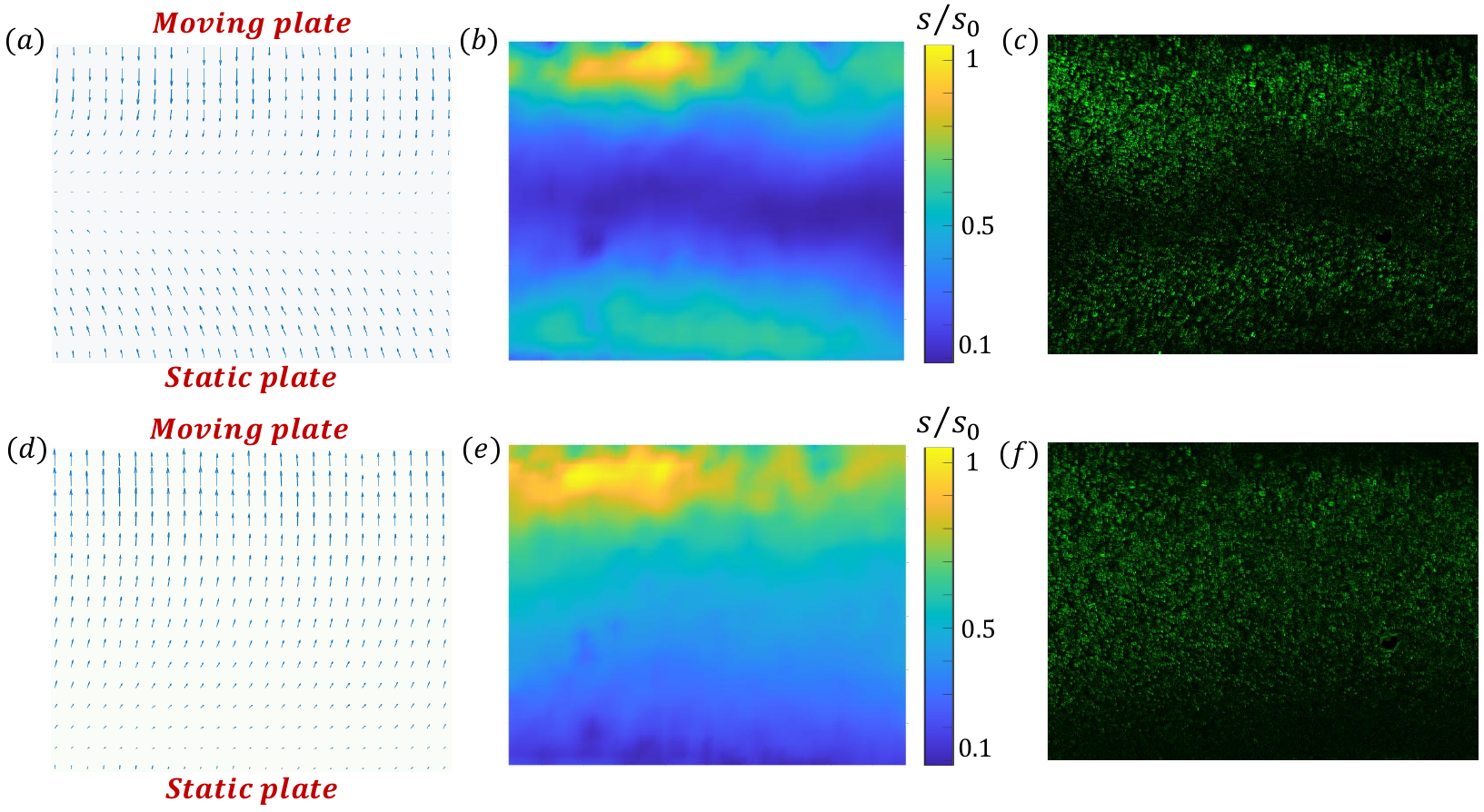}
 \caption{ (a) Displacement field shown as a vector map indicating inward movement (localized compression of the network) during stress relaxation of the first step strain deformation. The red arrow indicates the moving plate. The distribution of displacement magnitudes (panel b) and the image difference at the sample interface (panel c) during the first step show the prominent sample reorganization near both the static and moving plates. (d) Displacement field shown as a vector map indicating outward movement (localized expansion of the sample) during stress relaxation of the second step strain deformation. The distribution of displacement magnitudes (panel e) and the image difference at the sample interface (panel f) during the second step show the prominent sample reorganization near the moving plate.  }
\label{fig5}
\end{figure}

To gain further microscopic insights, we probe the local velocity profiles during rheology measurements. We perform in-situ boundary imaging at the sample-air interface to map the local deformation field \cite{mandal2025kovacs,bakshi2021strain,chattopadhyay2022effect}. Figure \ref{fig4}a presents the schematic of the in-situ imaging setup, where we capture the sample dynamics in the flow-gradient plane at the boundary of the rheometer plates. The imaging experiment is performed in a single drive configuration. To capture images at the sample boundary in this configuration, we have designed a bottom plate (static) with a diameter of 25 mm (same as the top cone geometry), which provides additional elevation of the bottom plate required for boundary imaging without losing coupling with the Peltier plate. Such customisation provides the flexibility to image the sample at a controlled temperature. As the collagen samples are transparent, we mix 1 wt\% of polystyrene tracer beads with a mean diameter of $\sim 3.3\, \mu m$ \cite{dhar2019signature} with the collagen sample before initiating the polymerization. Adding such a small amount of particles does not change the nature of the stress-strain curve and the observed Kovacs memory phenomenon, but provides enough intensity contrast in the boundary imaging to probe the sample dynamics using particle imaging velocimetry (PIV). Figure \ref{fig4}b shows a typical image of the sample-air interface seeded with polystyrene tracer beads. We analyze the images using PIV \cite{Thielicke_2021} to map the displacement/velocity fields under shear deformations. First, we map out the velocity profile under steady shear (with a strain rate of 0.01 $s^{-1}$) in both the linear and non-linear regions (also shown in Figure \ref{fig1}). Figure \ref{fig4}d shows the average velocity profiles at different strain values depicted as different colors. Here, the distance across the shear gap in the gradient direction is normalized with respect to the gap between the two plates at the boundary and the average velocity values at different distances are normalized by the velocity of the moving plate. The corresponding stress-strain plot is shown in Figure \ref{fig4}c, with data points marked using the same color scale as in the velocity profiles. We find that the velocity profiles remain linear in both the linear and nonlinear strain-stiffening regions of the system. However, beyond the strain-stiffening region, the system yields and undergoes plasticity and failure, as reflected in the nonlinear nature of the velocity profiles at large strain values. Similar yielding behaviour has been reported in the literature \cite{bakshi2021strain,burla2020connectivity}. Notably, although the stress-strain relationship differs significantly between the linear and the strain-stiffening region, the associated velocity profiles remain unchanged, showing affine deformations. As mentioned earlier, the distinct, affine non-linear regime well before yielding is a unique feature of biopolymer networks.
\newline
\newline
Next, we take a closer look at the local deformation field obtained from the PIV measurements under the step strain protocol employed to study Kovacs effect [Figure \ref{fig2}a]. We want to clarify that all the PIV data are analysed during stress relaxation just after the driving plate comes to a complete rest, after reaching a set value of strain magnitude. 
Figure \ref{fig5}a shows the displacement vector field obtained during the stress relaxation after the application of the first strain step of magnitude $\gamma_1$. We find that the sample close to both the moving and the stationary plates tends to move inward, showing a contractile deformation during the stress relaxation. The displacement magnitudes of the sample near the plate boundaries are significantly higher compared to the sample in the middle [Figure \ref{fig5}b]. We also look at the localised sample reorganisation by taking image differences of the sample interface. The image subtraction is performed between the images captured at the beginning of the stress relaxation and at a later time in the stress evolution. Figure \ref{fig5}c depicts the difference image with a time difference of 3 seconds. We observe high-intensity speckles in the difference image near the sample boundaries. This indicates that the reorganisations in the sample during the stress relaxation take place predominantly in the regions close to the sample boundaries. This observation is also consistent with Figures \ref{fig5}a and \ref{fig5}b. 
We next probe the velocity fields after the second step perturbation. We find that the portion of the sample starts to move towards the moving plate, indicating an overall tendency of the network to expand as shown in Figure \ref{fig5}d. The displacement magnitude shown in Figure \ref{fig5}e suggests that the displacement of the sample gradually increases as the moving plate is approached from the static one. This observation is consistent with the difference-image analysis, which shows enhanced sample reorganisation near the moving plate compared to the static plate (see Figure \ref{fig5}f). This suggests that the system undergoes a compressive deformation after the application of $\gamma_1$ due to a strong negative normal force in the system. As the sample gets gradually compressed, it slowly relaxes the negative normal stress [Figure \ref{fig3}a]. We find an overall expansion-dominated response after $\gamma_2$. This may result from the sudden decrease in the negative normal stress magnitude due to the application of $\gamma_2$: The compressed portion of the sample due to $\gamma_1$, starts to show a viscoelastic retraction due to a release of compressive stress [Figure \ref{fig3}a]. Remarkably, the regions of the sample that undergo spatial reorganisation during the application of the first and the second steps are different from each other. As we observe from Figures \ref{fig5}a, and \ref{fig5}b, the compression happens almost equally near the static and the moving plates. On the other hand, the expansion after $\gamma_2$ predominantly takes place close to the moving plate only [Figures \ref{fig5}d and \ref{fig5}e]. This indicates that the different relaxation modes required for observing the Kovacs-like memory effect can be spatially distinct. A feature that we report for the first time in the context of Kovacs effect. 

\section{Conclusion}

In conclusion, we report a robust Kovacs-like memory effect in a strain-stiffening biopolymer network, demonstrated here for the first time. This memory effect emerges exclusively within the nonlinear strain-stiffening regime. Notably, compared with previously reported studies on the Kovacs effect under mechanical perturbations, our system exhibits pronounced non-monotonic stress relaxation over a broad range of $\frac{\Delta \gamma}{\gamma_1}$ values. We believe that the high relaxation rates in the strain stiffening regime, even for longer waiting times, are responsible for observing non-monotonic relaxation for a wide parameter range. For other systems, where non-linearity comes due to material yielding, the relaxation rates slow down at longer waiting times. This highlights the critical role of strain-stiffening response and the associated enhancement of stress-relaxation rates in enabling the Kovacs effect across a wide spectrum of perturbations. However, additional studies are required to establish such a conclusion.

Because many colloidal gels also display nonlinear strain-stiffening responses, it will be compelling to investigate whether such colloidal systems can similarly exhibit a Kovacs effect over a comparably broad parameter space. As indicated by the phase diagram, except at very large values of $t_w$ and $\Delta \gamma$, the non-monotonic evolution of normal stress correlates strongly with the corresponding variations in shear stress. This observation underscores the central role of normal stress in governing memory phenomena in biopolymer networks.

Our in-situ boundary imaging experiments further reinforce this conclusion: the sample’s relaxation dynamics are predominantly driven by compressive or extensile motions. Strikingly, we also observe a spatial decoupling of relaxation modes between the first and second perturbation steps, with distinct regions of the sample undergoing reorganization at different stages. Identifying the fast and slow relaxation modes, along with their spatial distribution, an essential ingredient in understanding the Kovacs effect, remains an important avenue for future investigation.


%
%

\ack{We would like to thank MK Firoz for the useful discussion.}

\funding{SM acknowledges the intramural funding from Raman Research Institute for support.}

\roles{SM and AG conceived the project. AG performed the experiments and analysis. AG and SM wrote the manuscript.}

\data{The data that support the findings of this study are available from the corresponding author upon reasonable request.}

\suppdata{See the supplementary material for information about movie description and additional measurements. }

\bibliographystyle{iopart-num}
\bibliography{mybib}

\pagebreak

\begin{center}
\title{\Large Supplementary Information: Kovacs-like memory effect in strain stiffening collagen networks}    
\end{center}

\setcounter{section}{0}
\setcounter{figure}{0}
\setcounter{page}{1}

\renewcommand{\thefigure}{S\arabic{figure}}
\renewcommand{\thesection}{S\arabic{section}}

Access to supplementary movies: \url{https://drive.google.com/file/d/1MPiypm9NNoQ9okKMKCRD6vjJ_Pxfkuqc/view?usp=sharing}

\section{Movie Descriptions}
\subsection{Movie 1}
   This movie depicts the spatial reorganization of the sample interface during a two step Kovacs memory phenomenon. During the stress relaxation of the fist step strain, the sample shows contractile deformation depicted by the inward orientation in the vector field. However, the sample tend to expand depicted by the outward orientation in the vector field during stress evolution after the second step is applied. The vector field is mapped using PIV algorithm.

\vspace{6cm}
   
\section{Supplementary Figures}  

\begin{figure}[hbt!]
\centering
\includegraphics[height=7cm]{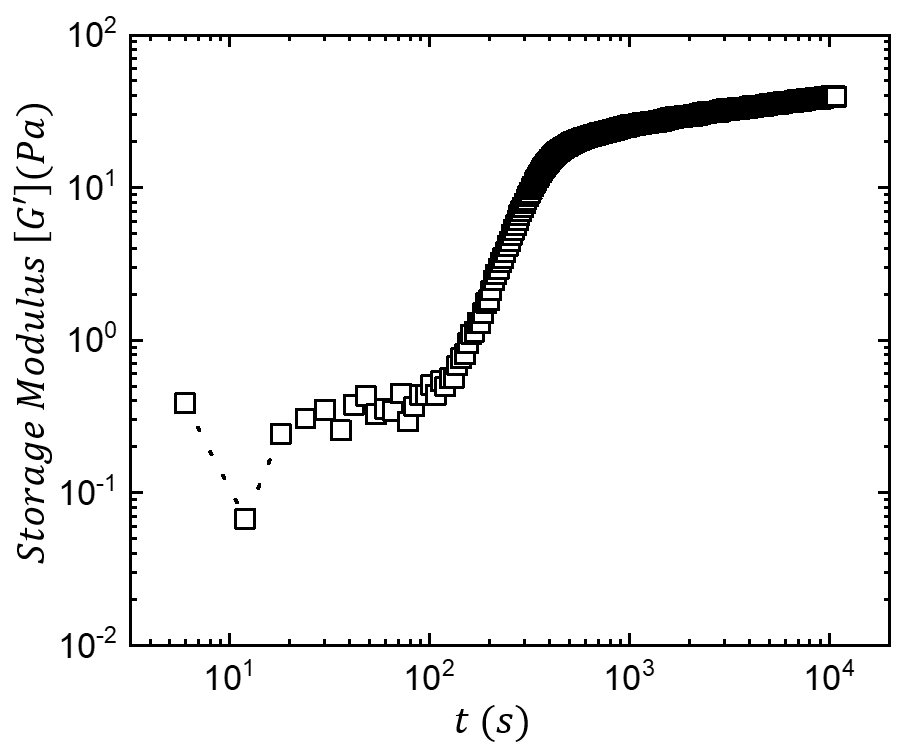}
\caption{\label{fig:wide} Variation of storage modulus ($G'$) as a function of time during polymerisation of collagen networks with 2 mg/ml monomer concentrations at a temperature of $25^{\circ}C$. The applied oscillatory strain amplitude is 2\% and frequency is 0.5 $Hz$.  At shorter times, $G'$ is smaller and, begins to increase and eventually saturates at longer times after the polymerisation process is complete. }

\end{figure}

\begin{figure}
\centering
\includegraphics[height=6.5cm]{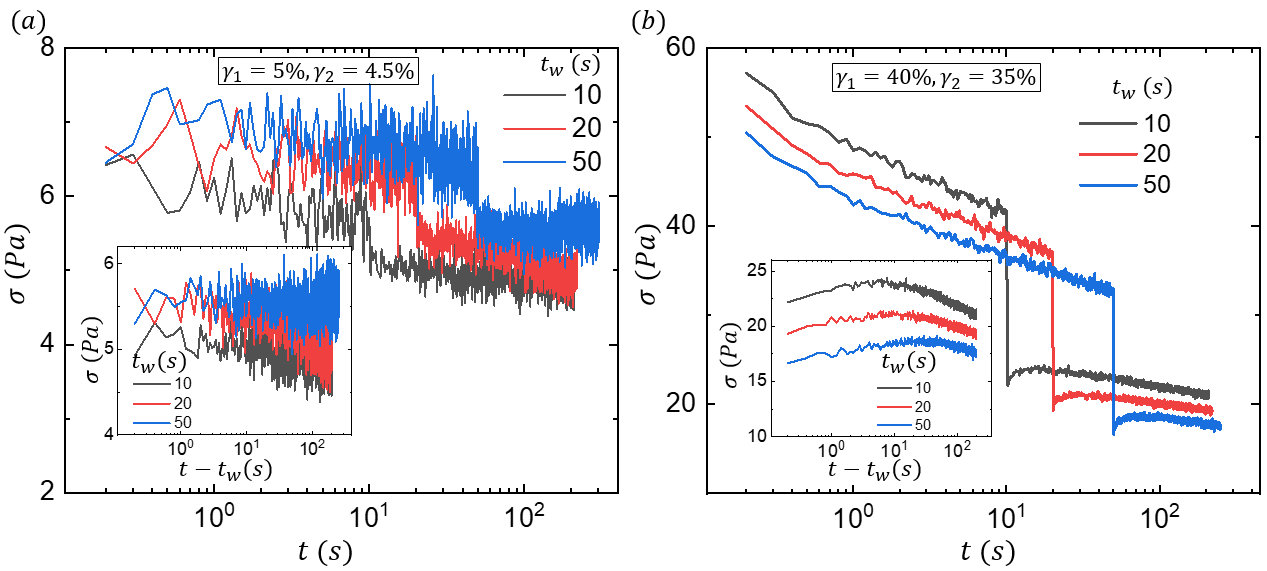}
\caption{\label{fig:wide} (a) Stress response during two step strain deformation for the Kovacs memory corresponding to $\gamma_1 = 5\%$ that lies in the linear region of the stress-strain response (see Figure 1 in the main text). The non-monotonicity is stress is not observed (also shown in the inset) and hence Kovacs memory is not found. (b) Stress response during two step Kovacs memory experiment for $\gamma_1 = 40\%$ that lies in the non-linear strain stiffening region (see Figure 1 in the main text) showing clear non-monotonic evolution.}

\end{figure}

\begin{figure}
\centering
\includegraphics[height=4.2cm]{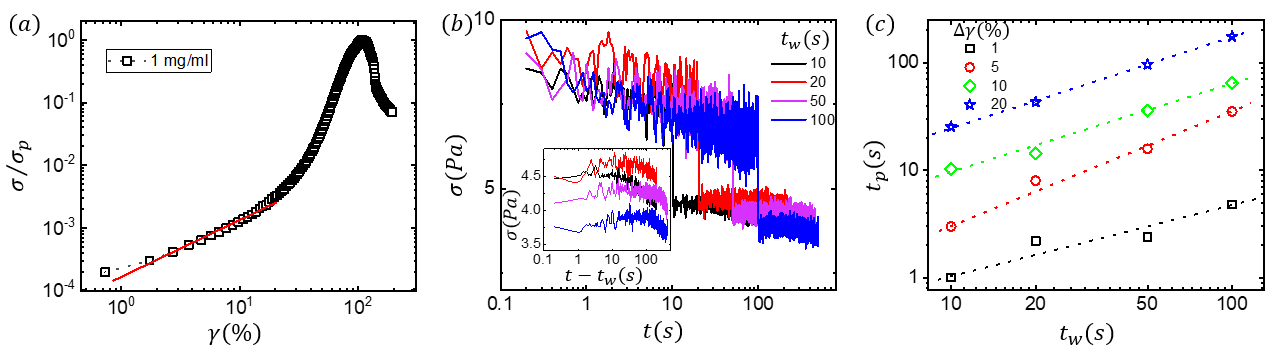}
\caption{\label{fig:wide} (a)Stress-strain curve for 1 mg/ml collagen polymerized at $25^{\circ} C$. The linear relation between stress and strain is marked as the red line. (b) Temporal evolution of stress showing Kovacs effect for $\gamma_1$=40\%, \, $\gamma_2$ = 35\%,  for different values of $t_w$ marked as different colors, and the inset shows the stress evolution during the second step $\gamma_2 = 35\%$. Peak is shear stress is observed in all the case. Relation between the peak time and the waiting time for $\gamma_1 = 40\,\%$ for different values of $\Delta\gamma$.}
\end{figure}

\begin{figure}
\centering
\includegraphics[height=4.2cm]{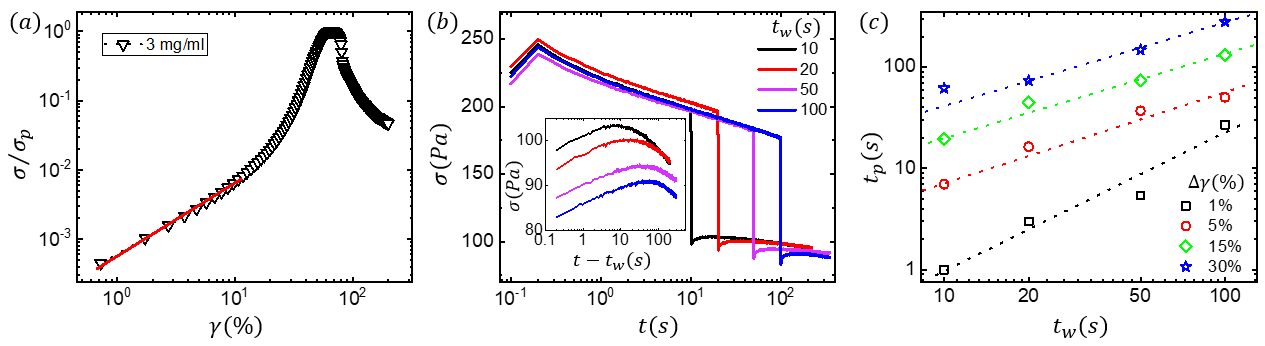}
\caption{\label{fig:wide} (a)Stress-strain curve for 3 mg/ml collagen polymerized at $25^{\circ} C$. The linear relation between stress and strain is marked as the red line. (b) Temporal evolution of stress showing Kovacs effect for $\gamma_1$=40\%, \, $\gamma_2$ = 35\%,  for different values of $t_w$ marked as different colors, and the inset shows the stress evolution during the second step $\gamma_2 = 35\%$. Peak is shear stress is observed in all the case. Relation between the peak time and the waiting time for $\gamma_1 = 40\,\%$ for different values of $\Delta\gamma$.}
\end{figure}

\begin{figure} [ht!]
\centering
\includegraphics[height=5.8cm]{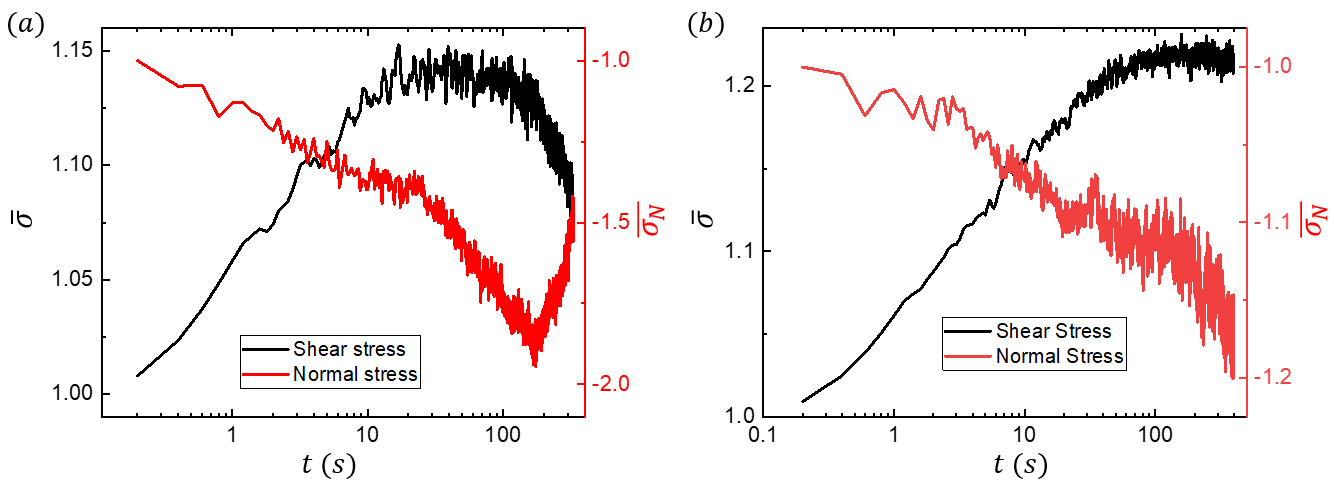}
\caption{\label{fig:wide} Time evolution of both shear and normal stresses (normalised w.r.t. first point) during the second step $\gamma_2$ for a two-step Kovacs memory protocol are shown in panel a and b. For panel a, $\gamma_1=40\%,\, \gamma_2 = 30\%,\,  t_w=50s $ and for panel b, $\gamma_1=40\%,\, \gamma_2=20\%,\, t_w=100s$. The peak time $t_p$ in shear stress and dip time $t_d$ in normal stress are not close to each other in panel a. In panel b, the non-monotonicity in shear stress is present, but absent in the case of normal stress. }

\end{figure}

\end{document}